# Can a photon be separated from its wave?

Louis Sica[1,2]

[1]*Institute for Quantum Studies, Chapman University, Orange, CA & Burtonsville, MD, 20866 USA*

[2]*Inspire Institute Inc., Alexandria, VA, USA*

Email: lousica@jhu.edu

An experiment is proposed in which the overall path taken by a photon is indicated by the timing of a twin herald photon, while a particular segment of that path is determined by interference. The needed coincident pairs of photons are generated by type I spontaneous-parametric-down-conversion (SPDC) and diffraction selected by a grating at the frequency of a two-photon state. This is to be divided into two one-photon states of high coherence length. Upon experimental confirmation that their count coincidences have been maintained, the one-photon sources are to be used as a timing herald and source for an unequal path interferometer. A photon's path through the interferometer via a short arm is indicated by count synchronization with the herald. The exit output port used and resulting final detection location are determined by the phase in the long arm. If output port usage can be controlled by the phase in the photon-free arm, the path of the photon as particle will have been controlled by interference with an accompanying photon-empty wave.

## I. Introduction

Among Feynman's more quoted statements is that the most fundamental quantum mystery [1] is that exhibited by the two-slit experiment and its demonstration of wave-particle duality. If two slit interference fringes are observed, how do these arise if light is composed of single particles as indicated by a detector's discrete counts? Is the photon a physical entity that can alter its nature from wave to particle in accord with the physical situation? Or is it possible that light is composed of both waves and associated particle-like entities coupled to the waves?

Several historic experiments have been performed that exhibit different aspects of wave-particle phenomena. Wheeler's Great Smokey Dragon experiment (performed in a recent version by V. Jacques et. al. [2]) shows that photon counts are anti-correlated after the first beam splitter in a Mach-Zhender interferometer, demonstrating that each photon takes one path or the other. Yet, in the same apparatus, detector signals obtained after the introduction of a final beam splitter indicate intensity variation that may be accounted for by wave interference due to path length differences for single photons (waves?) traversing the whole interferometer.

A more recent exhibition of the phenomenon in the Afshar experiment [3] employed an arrangement in which beams are formed from two pinholes followed by an imaging lens to send the light from each pinhole to a corresponding but different detector. Two beam interference fringes occurred in an intermediate region of wave overlap after which the beams separated and propagated to their individual detectors. A thin wire grating with wires placed in the fringe zero-intensity minima allowed the existence of fringes in the region of two-beam superposition to be monitored while counts from the two detectors remained anti-correlated. While for some, this experiment and its offshoots [4] seem to demonstrate that which-path information and interference can be obtained from the same set of events, others have found the experimental outcome insufficient to warrant their abandonment of the usual quantum interpretation [5].

In conventional quantum theory, these kinds of results, waves experimentally determined to go both ways, but photons traversing one path or the other, are accounted for by assuming that an individual photon is in a superposition state occupying both paths after a beam splitter. It then collapses to a single detector area as a result of measurement. Finally, in quantum electrodynamics, a photon is described as the first excited state of a lightwave filling all of space [6]. It may therefore be an entity of unlimited size that upon measurement instantaneously collapses to a detector of perhaps microscopic size, thereby exceeding the velocity of light. Many have found such an interpretation of quantum formalism to be inconsistent with relativity, though it leads to a correct numerical description of observations when employed in the statistical formalism. The question, however, is whether the formalism, under this interpretation, completely describes physical reality and is unique, or whether other, yet unrecognized physical processes may lead to more logically consistent explanations of such optical phenomena.

Experimental facts exist that suggest alternative approaches to explanations of problematic quantum phenomena. One of these is the existence of the Casimir effect [7] in which a photon-free electromagnetic field, the vacuum field, produces forces between objects. This indicates that the photon-less vacuum field predicted by the QED formalism is a real physical entity. Its energy is half of a photon's energy. Its reality is further suggested by the fact that atomic spontaneous emission rates in a cavity mode become very long [8] if the existence of the mode is prohibited by boundary conditions. This is consistent with the existence of an interacting photon-free vacuum wave mode that enables spontaneous emission.

Recently, pilot wave theory and experiment, described in [9], have suggested an alternative approach to understanding wave-particle duality. Reference [9] describes a phenomenon in which particles, bouncing liquid drops in this case, interact with waves in a fluid in such a way that particle and wave interference phenomena occur that are analogous to those of optics, although in a very different physical situation. Details of corresponding interactions of photons with a vacuum field are unknown to the author, but the fact that a concrete explanation of wave-particle duality exists for a macroscopic situation motivates the search for an analogous particle-coupled-to-wave explanation of optical phenomena such as those cited above. To be viable, many physical details of interaction would ultimately need to be provided. Further, experiments will need to be devised demonstrating new optical phenomena consistent with an expanded explanation of quantum optical phenomena.

The suggestion of two linked experiments given below is an effort in this direction, in spite of the fact that the phenomenon one is attempting to create has been believed to be disallowed by an accepted quantum mechanical interpretation. It has been thought impossible *in principle* to create a situation in which interference occurs if interferometer which-path information is known. In the proposed experiment, which path information is to be provided through timing and count synchronization using a twin-beam herald photon. If realizable, it would follow that interference due to the chosen phase of a wave in a photon-empty interferometer arm can determine the output port for a photon that has taken the other path to a final interferometer beam splitter. If observed, this outcome would be consistent with the idea that light consists of both particles and waves, rather than an entity that switches between the two. It would also be compatible with the Smokey-Dragon, Casimir, and Afshar experiments mentioned above.

It may be observed that the proposed experiment is analogous to that of the quantum Cheshire cat (see [10] and references cited within) in which a photon in one interferometer arm is interpreted as separated from its polarization in the other arm. The design of the Cheshire cat experiment derives from the concept of weak measurements [11] which has received considerable recent attention [10]. The analogous experiment suggested here is based on strong measurements with the photon path taken indicated by timing with the herald. It is thus related to "Welcher Weg" and quantum eraser experiments such as those described in [12,13].

## II. Action of a diffraction grating filter in a particle-coupled wave picture

The interferometer experiment described in Sec. IV requires a narrow-band optical filter to produce a long coherence length from the SPDC light that it transmits. To accomplish this in one pass, the use of a high efficiency (e.g. ~ 90%) diffraction grating is proposed. However, the effect of this component on down-converter synchronized photons must be subject to prior experimental examination [14]. The outcome depends on whether the closely degenerate collinear component of output of a Type I (SPDC) consists of initially localized photon pairs that remain localized in the process of grating diffraction, or whether the photons' two positions are dispersed over the wave-front in the process of diffraction. In the latter case, the lengths of their paths to associated detectors would be random over the ensemble of events resulting in a decrease in pair synchronization depending on grating width. The experiment would thus have implications regarding the nature of a photon, and whether local pilot wave concepts are qualitatively consistent with optical experiment.

However, there are two rather different situations to be considered. In the first, just indicated, one assumes a type I SPDC operated in a collinear, degenerate mode [15]. This is followed by a beam expander consisting of a negative followed by a positive lens so that the integrity of optical wave-fronts of different spectral components is maintained.

If a photon or photon pair of a given frequency exists locally, and was previously attached to a corresponding wave, it must be diffracted by a grating in a direction corresponding to that frequency. Consequently, it may be reasoned that wave-attached photon pairs *created at the same time and place* [15,16] in a down converter should maintain the same relative spatial position after diffraction as they had before (see Figure 1). If that is the case, their temporal correlation should be maintained.

An alternative conclusion follows if a quantum interpretation is used in which photons have no actual location before measurement since wave properties of light determine interaction with a diffraction grating. Different points on an input wave-front of a diffracted wave have different travel times to the output wave-front depending on their initial position on the wave-front (Figure 1). If the relative spatial positions of photons of a pair are defined only by the width of the wave-front before measurement, their travel time cannot be more precisely defined than by that of the wave-front as a whole. In that case, the temporal coincidence of degenerate, collinear photons would not be expected to be maintained.

Even if the locality of photon pairs is maintained, however, a question arises as to how a grating monochromator decreases the bandwidth of the light. When the bandwidth is decreased and the wave train lengthened, do two paired photons of almost the same frequency due to phase matching maintain their relative longitudinal positions (and ultimately their count synchronization) after diffraction?

Due to the large bandwidth produced in the down-converter process, a large width grating would be necessary to produce the long coherence length needed in the second part of the experiment if single photon pairs are used. If count synchronization fails to be maintained, the second experiment could not be carried out. If count synchronization is maintained, however, the large width grating needed under this scenario could make the overall experiment challenging.

However, results of two recent experimental investigations [17,18] of SPDC output indicate properties that appear to greatly simplify the current proposal. In Reference [17], it was found that a coincidence count time window of 330 ps, produced a Michelson interferometer fringe visibility of 0.87 with an interferometer path difference of 27 cm. This was accounted for given the selection of a two-photon state count-synchronization based on the narrow time window used and the overall experimental setup. In addition, the fringe spacing resulting from interferometer scanning indicated a two-photon state interference at half the spacing of interference of two one-photon states.

In reference [18], the light from a down converter was passed through a coarse transmission grating, and it was found that the spatial distribution of two-count coincidences corresponded to well defined diffraction orders at the angular position of the two-photon wave frequency i.e. the pump frequency, rather than that of two one-photon waves at half the pump frequency. The two photon wave was later converted to two synchronized one-photon waves at half the pump frequency by a beam splitter.

## III. Expected coherence length from a grating in two cases

The light from the collection of one-photon pair-states produced by a SPDC has a very short coherence length (16 $\mu m$ after spectral filtering in Ref. [19]), too short to allow efficient functioning of the interferometer experiment proposed in Sec. IV. The coherence time $T_c$ is the reciprocal of the bandwidth $d\nu$, so that the coherence length is $l_c = c/\Delta\nu$. Because the coherence length is small, the visibility of fringes in the interferometer would not be sufficient to produce observable/efficient switching between detectors D1 and D2 as described in Sec. IV. Thus, to realize the experimental test that is proposed herein, output radiation of the SPDC requires spectral filtering in a way that maintains count synchronization originating in photon pairs. An approach to accomplishing this would be to use a sufficiently wide diffraction grating of high efficiency assuming that one-photon pair-states are to be used for the experiment.

The diffraction grating width needed may be estimated by the requirement to produce a sufficiently large path difference in the interferometer, so that the time delay for photons traveling one path versus the other is measureable. This needs to be accomplished with a high fringe visibility so as to produce efficient switching between detectors D1 and D2 as described in Sec. IV. The grating width may be computed from [20]

$$\frac{\lambda}{d\lambda} = \frac{w(\sin\theta - \sin\theta_o)}{\lambda}\ , \tag{1}$$

where w is the width of the grating, $\theta$ and $\theta_o$ are angles between the beam directions and the grating normal, and $d\lambda$ is the spectral width/resolution of the beam at $\theta$ and $\theta_o$. Since the coherence length is c times the coherence time $T_c \approx 1/dv$, one finds from Equation (1) that

$$l_c = cT_c = c/d\nu = c(\lambda^2/cd\lambda) = w(\sin\theta - \sin\theta_o)\ . \tag{2}$$

Thus, the path difference in the interferometer should be considerably smaller than the path difference between the beam edges at the diffraction grating to produce high visibility interference, and yet be large enough to allow distinguishing photons that traverse the short path versus the long path by time synchronization measurements. (The spectral filtering effect of the grating on the two assumed photons is indicated in Fig. 2 by the words "Spectral filters".)

The above comments apply to the situation assuming paired one-photon states are to be used in the experiment. However, in [18] it was shown that a grating could select two-photon states from a SPDC Type I output in place of the two one-photon states just considered. This appears to have great advantages both with respect to achieving a narrow bandwidth (long coherence length) and high count synchronization. The selected two-photon states have a wavelength equal to that of the pump but are orthogonally polarized to it. The one-photon states have wavelengths of the order of twice the pump wavelength. The grating width needed to perform the two-photon state selection may thus be much smaller than indicated by Eq. (2) applied to one-photon states.

The wavelength difference between the two-photon and one-photon states is

$$\left|\lambda_p - 2\lambda_p\right| = \lambda_p \, . \tag{3}$$

Using Eq. 1 above in a configuration with $\sin\theta - \sin\theta_0 \cong 1$,

$$\frac{\lambda_p}{d\lambda_{passbamd}} = \frac{\lambda_p}{\lambda_p^2 / w} = \frac{w}{\lambda_p} \tag{4}$$

For a 1 cm width grating and a pump of 0.5 micron, the wavelength difference divided by the grating pass-band at the pump frequency is $2 \bullet 10^4$ . For a one millimeter grating the number is 2000. Thus a beam expander may not be needed to obtain the necessary wavelength separation.

The separation of two-photon states from the wide down-converter spectrum followed by their conversion to one-photon states for final use as in Ref. [18] could greatly facilitate performance of the experiment proposed in Sec. IV. This is because the bandwidth of the resulting one-photon states would be expected to be determined by that of the laser, as opposed to the large bandwidth of one-photon states not derived from two-photon states. (Note: one-photon state photons would be eliminated by using only synchronized counts.) Using this second strategy, the experiment of Sec. IV appears to be considerably more feasible than using the first strategy.

## IV. Can which-path information and wave interference exist simultaneously?

The interferometer and its function will be described with reference to the schematic in Figure 2. The experiment uses a Michelson interferometer with a non-polarizing, symmetrical 50-50 beam splitter, and return mirrors replaced by corner reflectors. The light source for the experiment is a Type I SPDC adjusted so that the predominant emission consists of photon pairs with the pump removed by a polarization beam splitter [18]. The photon emitted on the left serves as a timer for the second photon emitted toward the interferometer. The SPDC output is spectrally filtered using the technique described in Sec.III to select photons emitted at the same time with the same frequency. (In the scheme described above, half of photon pairs will both go to the herald side or the interferometer side and will not be counted.)

The corner reflectors of the interferometer enable two beam outputs at P5 to be sent to two separate detectors, and it is assumed that the interferometer is adjusted so that beam phase is spatially constant over the beam diameter. The two detectors D1 and D2 are equidistant from the output beam splitter at P5. Under the symmetrical conditions postulated, a maximum output at D1 corresponds to a null at D2 and vice versa as a function of the path difference $\Delta L$ between the two arms of the interferometer.

The interferometer is adjusted in the Figure so that path P4P5 via R1 is shorter than P4P5 via R2. Further, the path P3P4R1P5D1 equals path P3P4R1P5D2. These paths equal P2P1 in traversal time so that photons that travel the short path through the interferometer will produce coincidence counts between D3 and either D1 or D2. However, if photons travel through the interferometer via longer path R2, their counts will not be in coincidence with D3 but will occur after a predictable time delay. Thus, for coincidence counts between either D1 or D2, and D3, the photon must have taken the short path via R1. The final requirement of interferometer function is that for small variations in the longitudinal position of R2, the beam at P5 from R2 can be varied in phase by $\pi$. Due to these variations in phase and their effect on beam interference, the output power (photon counts) may be switched between D1 and D2. (It may be observed that by readjustment of path-length P2P1 to match the length of the long interferometer arm, short arm re-positioning of R1 could be used to implement detector switching. One could also observe photons taking the long path by taking the time delay into account.)

The fundamental question then is this: if one observes coincidences between the herald photon at D3, and D1 or D2, can these coincidences be switched between all-D1 and all-D2 (assuming proper adjustment of the interferometer), by quarter wave position variations of R2, thus varying the long path phase at P5? If that is the case, then in spite of having which-path information for the photon through the interferometer, interference will still have been observed.

## V. Discussion and Conclusion

Recent experiments such as [17,18] show that a two-photon state may be formed in the degenerate co-linear output of a Type I SPDC, and that its position is distinguishable in diffraction grating output from the wideband radiation consisting of pairs of one-photon states also generated by the nonlinear process. If the two-photon state is divided into two equal frequency one-photon states at a beam splitter, as suggested by experiments and theory, the bandwidth of the resulting photons should be very narrow and their synchronization very high. The use of the photons as herald and source for an unequal path interferometer would then seem to be possible. If the bandwidth of these one-photon states is not narrow with high count synchronization, however, the interferometry experiment would not then be feasible. Thus, the experimental effort naturally divides into two parts.

If pairs of single photons at twice the pump wavelength are used to form the herald and interferometer source, traveling closely similar paths from a common location of creation, their coincidence rate should not be greatly effected by grating diffraction. Their paths determine their coincidence times in a guided wave picture. However, if photon particle-like behavior manifests itself only upon observation induced wave-function collapse, then the notion of a photon attached to a wave, as in the pilot wave model, will be unsupported by experimental count-synchronization statistics. Comparisons of photon coincidences before and after grating reflection should provide evidence regarding this issue. As before, the outcome will determine the feasibility of conducting the second part of the experiment. However, if single photons are derived from pairs of one-photon states to produce the herald and interferometer source, a much larger grating will be needed to obtain the required coherence length for the interferometer experiment. The coherence length resulting from the spectral filtering must be long enough to produce high visibility interference at P5. In addition, photon coincidences using a narrow time window are needed to produce unambiguous and efficient photon switching. Finally, of the two approaches described, the one beginning with a two-photon state at the pump frequency appears to be the simpler.

The suggested experiment might be considered to be an extension of the "Smokey Dragon" experiment in which single photons show interference effects upon traversing an interferometer. If twin photons can be produced at the same time with narrow spectral band width, it would appear possible to use one as a herald for the other traversing an unequal path interferometer. Since the one traversing the interferometer would not be disturbed, it would be expected that interference could be observed as in the "Smokey Dragon" experiment.

**Acknowledgements**

I would like to thank Mike Steiner for useful discussions at an early stage of development of the idea, as well as critical comments on the manuscript. I would also like to thank Armen Gulian and Joe Foreman of the Quantum Studies Group at Chapman University, Burtonsville, MD for useful discussions, and particularly John Reintjes of Sotera Defemse Solutions for an incisive critique of the manuscript that motivated several changes.

## References

[1] R. P. Feynman, R. B. Leighton, and M. Sands, *The Feynman Lectures on Physics, Vol. 3*, (Addison-Wesley, Boston, Mass, 1965). pp. 1.1–1.8. ISBN 0201021188.

[2] V. Jacques, E. Wu, F Grosshans, F. Treussart, P. Grangier, A. Aspect, and J. F. Roch. Science 315, 5814, (2007) 966 arXiv:quant-ph/0610241

[3] S. S. Afshar, *The Nature of Light: What Is a Photon?* Proc. SPIE. **5866** (2005). p. 229. arXiv:quant-ph/0701027. doi:10.1117/12.638774

[4] S. S. Afshar; E. Flores; K. F. McDonald, and E. Knoesel, Found. Phys. **37**(2), 295 (2007). arXiv:quant-ph/0702188. doi:10.1007/s10701-006-9102-8.

[5] O. Steuernagel, Found. Phys. **37** (9), 1370 (2007). arXiv:quant-ph/0512123. doi:10.1007/s10701-007-9153-5.

[6] R. Loudon, *The Quantum Theory of Light*, (Oxford University Press Inc., New York, Third Edition, 2000). Chap. 4.

[7] H. G. B. Casimir, Proc. K. Ned. Akad. Wet. 51, 793 (1948).

[8] D. Kleppner, Phys. Rev. Lett. **47**(4), 233 (1981). doi:10.1103/PhysRevLett.47.233.

[9] J. W. M. Bush, Phy. Today, 68(8), 47 (2015).

[10] J. M. Ashby, P. D. Schwarz, and M. Schlosshauer, Phys. Rev. A **94**, 012102 (2016), DOI: 10.1103/PhysRevA.94.012102

[11] Y. Aharonov, D. Z. Albert, and L. Vaidman, Phys. Rev. Lett. **60**, 1351 (1988).

[12] M. O. Scully and K. Druhl, Phys. Rev. A **25**, 2208 (1982).

[13] Yoon-Ho Kim, R. Yu, S. P. Kulik, Y. H. and Shih, M. O. Scully, Phys. Rev. Lett. **84**, 1 (2000).

[14] M. Bashkansky, J. Reintjes, M. Steiner, personal communications.

[15] Y. Shih, *An Introduction to Quantum Optics*, (Taylor & Francis, Boca Raton, FL, 2011). pp. 296-297.

[16] G. Grynberg, A. Aspect, and C. Fabre, *Introduction to Quantum Optics*, (Cambridge University Press, Cambridge, New York, 2010). p. 552.

[17] J. Brendel, E. Mohler, and W. Martienssen, Phys. Rev. Lett. 66(9), 1142 (1991).

[18] R. Shimizu, K. Edamatsu, and T. Itoh, Phys. Rev. A 74, 013801-1 (2006).

[19] C. K. Hong, Z. Y. Ou, and L. Mandel, Phys. Rev. Lett. 59(18), 2044 (1987).

[20] M. Born and E. Wolf, *Principles of Optics*, (Cambridge University Press, New York, NY, Seventh edition 1999). p. 452.

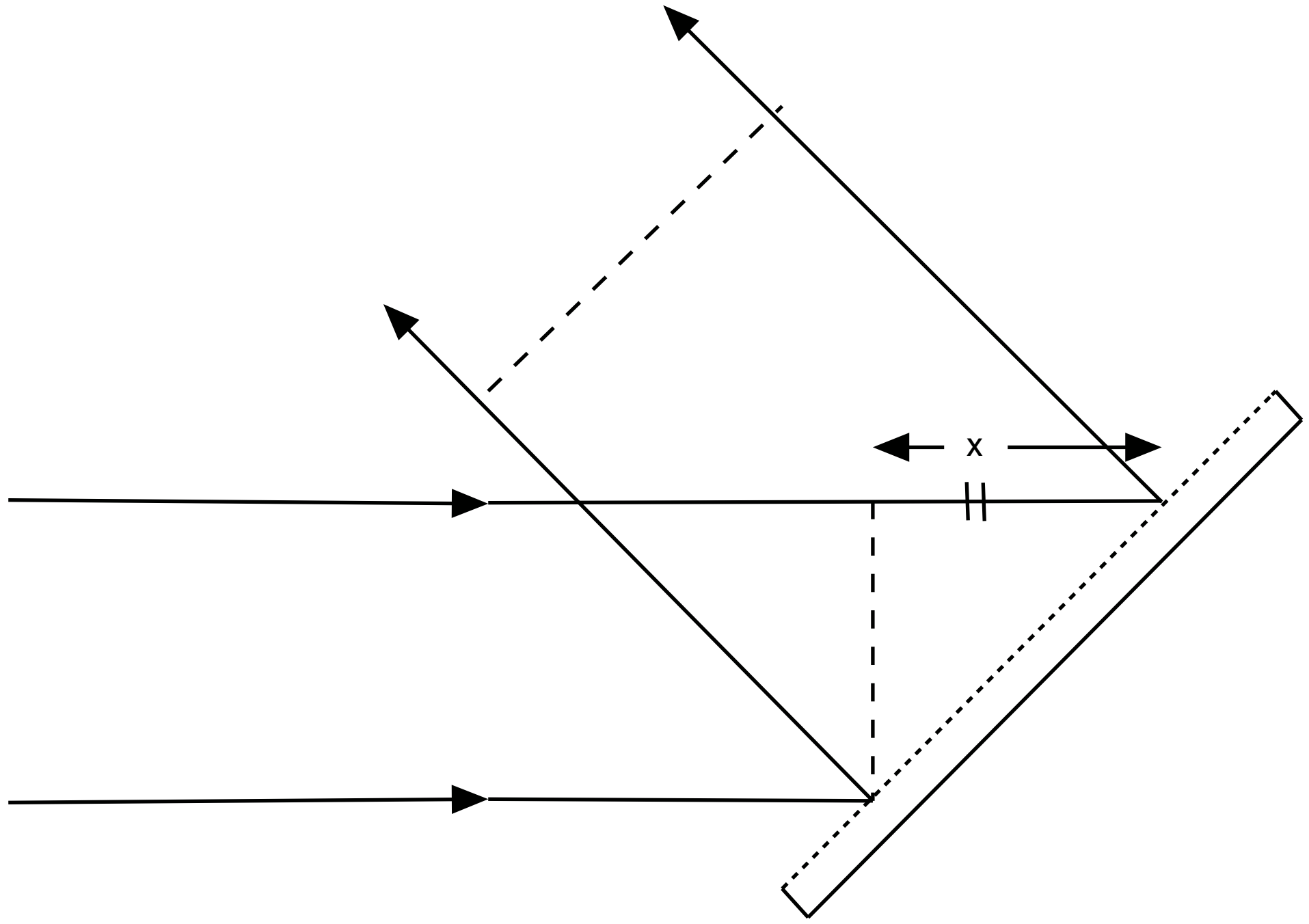

FIG. 1. Parallel beam diffracted in the direction of the normal to a reflection grating. The path length difference of a ray at the top of the beam relative to the bottom of the beam is x. For other rays, it varies with position along the wavefront.

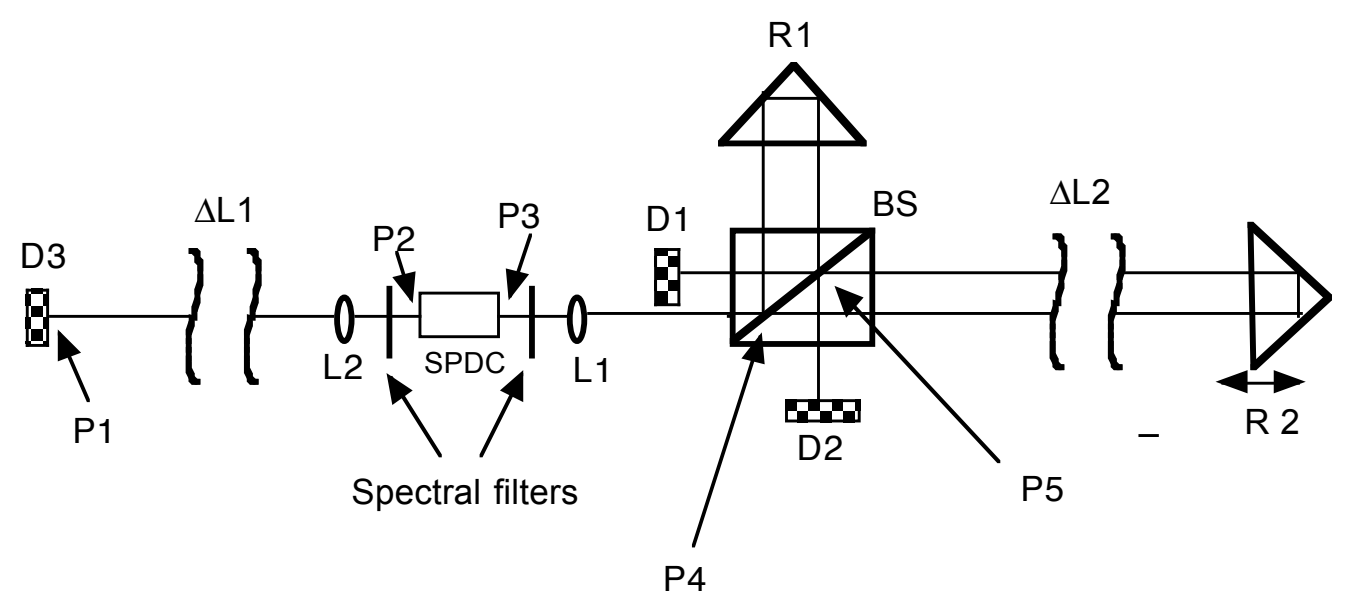

FIG. 2. Schematic of optical system for switching output port by means of phase shift in the unused arm of an interferometer. The function of the two spectral filters schematically shown is to be realized using the monochromator-like arrangement described in the text.